\documentclass[twocolumn,showpacs,preprintnumbers,amsmath,amssymb]{revtex4}

\usepackage{graphicx}
\usepackage{dcolumn}
\usepackage{bm}

\begin{document}

\title{Strong limits on the possible decay of the vacuum energy into CDM or CMB photons}

\author{Reuven Opher}
\email{opher@astro.iag.usp.br}
\author{Ana Pelinson}
\email{anapel@astro.iag.usp.br} \affiliation{IAG, Universidade de
S\~{a}o Paulo, Rua do Mat\~{a}o, 1226 \\
Cidade Universit\'aria, CEP 05508-900. S\~{a}o Paulo, S.P.,
Brazil}

\date{\today}

\begin{abstract}
We investigate models that suggest that the vacuum energy decays into cold dark
matter (CDM) and into a homogeneous distribution of a thermalized cosmic microwave
background (CMB), which is characteristic of an adiabatic vacuum energy decay into
photons. We show that the density fluctuation spectrum obtained from the CMB and
galaxy distribution data agreement put strong limits on the rate of decay of the
vacuum energy. A vacuum energy decaying into CDM increases its total density $\rho$,
diluting $(\delta \rho/\rho)^2$. The observed temperature fluctuations of the CMB
photons $(\delta T/T)^2$ are approximately proportional to CDM density fluctuations
$(\delta \rho/\rho)^2$. In both case, when evaluating $(\delta \rho/\rho)^2$ at the
recombination era, its present measured value must be increased by a factor $F$.
Since the $(\delta \rho/\rho)^2$ derived from the CMB and galaxy distribution data
agree to $\sim 10\%$, the maximum value for $F$ is $F_{\rm max}\cong 1.1$. Our
results indicate that the rate of the decay of the vacuum energy into CDM and CMB
photons are extremely small.
\end{abstract}

\maketitle

\section{Introduction}

Recent observations has indicated that the universe is spatially
flat and undergoing a late time acceleration. This acceleration
has been attributed to a dark energy component with negative
pressure which can induce repulsive gravity. The simplest and most
obvious candidate for this dark energy is the cosmological
constant $\Lambda$ with the equation of state $w={p / \rho}=-1$,
where $p$ is the pressure and $\rho$ is the energy density. A
decaying vacuum energy is very attractive since it may link the
present vacuum energy that is accelerating the universe today with
perhaps the large vacuum energy that created the inflation epoch
in the past.

We analyze how the observed cosmic microwave background (CMB) and large galaxy
survey data constraints the decaying vacuum energy models into cold dark matter
(CDM) varying with redshift from the recombination era ($z\sim 1070$) to the present
($z\sim 0$) \cite{vdecay}. A vacuum energy decaying into CDM increases its total
density, diluting $(\delta\rho/\rho)^2$. In order to evaluate $(\delta \rho/\rho)^2$
at the recombination era, when it created the $\delta T/T$ of the CMB, its present
measured value obtained from the galaxy distribution data must then be increased by
a factor $F$. The density fluctuations derived from the CMB data were compared with
those derived from the 2dF Galaxy Redshift Survey (2dFGRS) \citep{2df,2df1}. Since
the $(\delta \rho/\rho)^2$ derived from the CMB and galaxy distribution data agree
to $\sim 10$ per cent, the maximum value for $F$ is $F_{\rm {max}}\cong 1.1$ (see
\cite{2dfend} for the final data set of the 2dFGRS).

We made a similar analysis for the possibility of the decay of the
vacuum energy into CMB photons \cite{cmbphotons}. In this
scenario, the temperature fluctuations at the recombination epoch
$( {\delta T}/{T})_{\rm rec}$ were diluted by the photons created
by the vacuum energy decay. Thus, the temperature fluctuations at
present are smaller than those existing at the recombination era.
This implies bigger density fluctuations $\delta\rho/\rho$ at the
recombination era than those derived from the observed CMB data.

\section{Vacuum energy decaying into CDM}

A vacuum energy decaying into CDM increases its total density,
diluting $(\delta\rho/\rho)^2$. Consequently, a larger density
fluctuation spectrum $(\delta\rho/\rho)^2$ is predicted at the
recombination era ($z_{\rm rec}=1070$) by the factor

\begin{equation}
F\equiv \left[ \frac{\overline{\rho}_M
\,(z)}{\overline{\rho}_M\,(z)-\Delta \rho (z)}\right]^{2}
\,\Big|_{z=z_{\rm rec}}\, \,, \label{fatorf}
\end{equation}
where
\begin{equation}
\overline{\rho}_M\,(z)=\rho _{c}^{0}\,\left(
1+z\right)^{3}\Omega_{M}^0  \label{rhostand}
\end{equation}
is the matter density for a constant vacuum energy density,
$\rho_c ^0 \equiv 3\,H_{0}^{2} / (8\,\pi \,G) \simeq
1.88\,h_{0}^{2}\times 10^{-29}\, g\,cm^{-3}$ is the critical
density, and $\Omega_{M}^0$ is the normalized matter density,
$\Omega_{M}^0=\rho^0_{M}/\rho^0_{c}$ ($\sim 0.3$).

The difference between the matter density $\bar{\rho}_M$ and the
matter density predicted by the model in which the vacuum energy
decays into matter, $\rho_{Mv}$, is
\begin{equation}
\Delta \rho(z) =\overline{\rho}_M(z)-{\rho}_{Mv}(z)\,.
\label{deltarho1}
\end{equation}

The density $\rho_{Mv}(z)$ is normalized at redshift $z=0$
$\left[\rho_{Mv}(z=0)\equiv{\rho^0}_{M}\right]$. In order to
describe the transfer of the vacuum energy $\rho_{\Lambda}$ into
matter $\rho_{Mv}$ \cite{bronstein}, we use the conservation of
energy equation,
\begin{equation}
\dot{\rho}_{\Lambda}+\dot{\rho}_{Mv}+3\,H\,(\rho_{Mv}+P_{Mv})=0\,,
\label{mequat}
\end{equation}
where $P_{Mv}$ is the pressure due to $\rho_{Mv}$. For CDM, we
have $P_{Mv}=0$.

There exists an extensive list of phenomenological $\Lambda$-decay
laws. Several models in the literature are described by a power
law dependence
\begin{equation}
\rho_{\Lambda}(z)= \rho_{\,\Lambda}^{\,0}\,(1+z)^{\,n}\,, \label{mainL}
\end{equation}
where $\rho_{\,\Lambda}^{\,0}\equiv \rho_{\Lambda}(z=0)$, which we
investigate here.

The solution for the matter density has the form
\cite{bronstein,peeb}
\begin{equation}
\rho_{Mv}(z)=A\,(1+z)^3+B\,\rho_{\Lambda}(z)\,,
\label{matter}
\end{equation}
where $A$ and $B$ are  unknown constants. Using Eqs.(\ref{matter})
and (\ref{mainL}) in Eq.(\ref{mequat}), the dependence of
$\rho_{Mv}$ as a function of $n$ is
\begin{equation}
\rho_{Mv}(z)=\rho_{Mv}^0 (1+z)^3- \frac{n\,\rho_{\Lambda}^0}{3-n}
\, \left[\,(1+z)^3-(1+z)^n\right]\,. \,\label{rhopeeb}
\end{equation}
Using Eqs.(\ref{rhostand}) and (\ref{rhopeeb}) in
Eq.(\ref{deltarho1}), we find from Eq.(\ref{fatorf}) that
\begin{equation}
F=\left[{1-\left( \frac{n}{3-n} \right)\,
\left(\frac{\rho_{\Lambda}^0}{\rho_{Mv}^0}\right)\,
\left[\,1-(1+z)^{n-3}\right]}\right]^{-2}\,. \label{facpeeb}
\end{equation}

If, as discussed in section I, the density power spectrum from observations can be
increased by no more than approximately $10\%$ due to the decay of the vacuum
energy, we then have a maximum value for the $F$ factor $F_{\rm max}\cong 1.1$. This
maximum value gives $n_{\rm max}\cong 0.06$.

We also considered a recent model suggested by the renormalization
group equation of the effective quantum field theory which has a
$\Lambda$-decay dependence \cite{shapiro}
\begin{equation}
\rho_{\Lambda}(z;\nu)=\rho_{\Lambda}^0+\rho_c^0\,\,f(z,\nu)\,,
\label{Lnu}
\end{equation}
where $\rho_{\Lambda}(z=0)\equiv \rho_{\Lambda}^0$, $k=0$, and
\begin{eqnarray}
f(z)=\frac{\nu}{1-\nu}\,\left[\left(1+z\right)^{3(1-\nu)}-1\right]\,.
\label{funcz}
\end{eqnarray}
The dimensionless parameter $\nu$ comes from the renormalization
group
\begin{equation}
\nu\equiv \frac{\sigma}{12\,\pi}\,\frac{M^2}{M_P^2}\,, \label{nu}
\end{equation}
where $\,\sigma M^2$ is the sum of all existing particles
(fermions with $\sigma=-1$ and bosons with $\sigma=+1$). The range
of $\nu$ is $\nu \in (0,1)$  \cite{shz}.

Using Eqs.(\ref{Lnu}) and (\ref{funcz}), the matter density as a
function of $z$ and $\nu$, in the matter era, is
\begin{equation}
\rho_{Mv}(z;\nu) \,=\,\rho_{Mv}^0 \,(1+z)^{3(1-\nu)}\,.
\label{Rnu}
\end{equation}
Using Eqs.(\ref{Rnu}) and (\ref{rhostand}) in
Eq.(\ref{deltarho1}), we find from Eq.(\ref{fatorf}), the factor
$F$ modifying the density power spectrum:
\begin{equation}
F=(1+z_{\rm rec})^{6\nu} \,. \label{runcos}
\end{equation}

Using $F_{\rm max}\cong 1.1$ we place an upper limit on the $\nu$ parameter:
$\nu_{\rm max}\cong 2.3\times 10^{-3}$.


\section{Vacuum energy decaying into CMB photons}

According to the standard model, the temperature fluctuations
observed today are given by the expression
\begin{equation}
\left(\frac{\delta T}{T}\right)\,\Big|_{z\sim {0}}\,= {\cal
K}\,\,\,\frac{\delta\rho}{\rho}\,\Big|_{{{z}_{\rm{rec}}}}\,,
\label{zto1070}
\end{equation}
where ${\cal K}$ is approximately constant and the temperature
dependence of $T(z)$ is
\begin{equation}
{T}(z)=T_0\left(1+z\right)\,, \label{tempstan2}
\end{equation}
where $T_0\simeq 2.75\,\rm{K}$ is the present CMB temperature
\cite{padm}. The present value of $(\delta \rho/\rho)^2$ is gotten
from the relation
\begin{equation}
\left(\frac{\delta\rho}{\rho}\right)\,\Big|_{z\sim {0}}\,={\cal
D}\,({z}_{\rm rec}\rightarrow z=0)
\,\,\,\frac{\delta\rho}{\rho}\,\Big|_{{z}_{\rm rec}}\,,
\label{phizto1070}
\end{equation}
where ${\cal D}\,({{z}_{\rm rec}}\rightarrow z=0)$ is the growth factor from the
recombination era until the present time.

When we assume that the decay is adiabatic, the vacuum energy
decays into a homogeneous distribution of thermalized black body
CMB photons and the standard linear temperature dependence becomes
modified \cite{Lima1}. The decay can be described by a generic
temperature dependence,
\begin{equation}
T(z)=T_0\,(1+z)^{1-\beta}\,, \label{tempbeta}
\end{equation}
of the CMB photons. In principle, the possible range of $\beta$ is
$\beta \in [0,1]$ \cite{Lima1}.

There are two effects due to the vacuum energy decaying into CMB
photons:
\begin{itemize}
\item [1)] Since the temperature fluctuations at the recombination epoch $(
{\delta T}/{T})_{\rm rec}$ should be diluted by the photons
created, the temperature fluctuations at present become smaller
than those existing at the recombination era.

\item [2)] The value of the recombination redshift ${\bar{z}_{\rm rec}}$ is
higher than that of the standard model ${{z}_{\rm rec}}$ since the
universe is cooler at any given redshift.

\end{itemize}

Due to the dilution of $\delta T/T,$ instead of Eq.(\ref{zto1070})
of the standard model, we must use the relation
\begin{equation}
F_1\,\left(\frac{\delta T}{T}\right)\,\Big|_{{{z}_{\rm rec}}}\,=
{\cal K}\,\,\frac{\delta\rho}{\rho}\,\Big|_{{{z}_{\rm
rec}}}\,,\label{first}
\end{equation}
where $F_1$ is defined by
\begin{equation}
F_{1}(z)\equiv \left[ \frac{{T}\,(z)}{{T}%
\,(z)-\Delta T(z)}\right] \,\Big|_{{{z}_{\rm rec}}}\, \,.
\label{factort}
\end{equation}
$\Delta T(z)$ is the difference between the recombination
temperature ${T} ({z}_{\rm rec})$ predicted by the standard model
and that of the model in which the vacuum energy decays into
photons at temperature $\overline{T} ({z}_{\rm rec})$:
\begin{equation}
\Delta T({{z}_{\rm rec}})={T}\,({{z}_{\rm rec}})-\overline{T}
({z}_{\rm rec})\,. \label{deltat2}
\end{equation}
Using Eqs.(\ref{tempbeta}), (\ref{factort}), and (\ref{deltat2}),
we obtain
\begin{equation}
F_1=\left(1+{{z}_{\rm rec}}\right)^{\beta}\label{f1beta}\,.
\end{equation}

From Eqs.(\ref{tempbeta}) and (\ref{deltat2}), $\overline{T} (z)$
was lower than ${T}\,(z)$ by $\Delta T$ at ${z}_{\rm rec}$. Thus,
the resultant recombination redshift $\bar{z}_{\rm rec}$ was
higher than that of the standard model ${{z}_{\rm rec}}$. Instead
of Eq.(\ref{phizto1070}), $\left({\delta\rho}/{\rho}\right)\,$ at
${z\sim {0}}\,$ is now given by
\begin{equation}
\left(\frac{\delta\rho}{\rho}\right)\,\Big|_{z\sim {0}}\,={\cal
D}\,(\bar{z}_{\rm rec}\rightarrow z=0)
\,\,\,\frac{\delta\rho}{\rho}\,\Big|_{z=\bar{z}_{\rm rec}}\,,
\label{phiztorec}
\end{equation}
where ${\cal D}\,(\bar{z}_{\rm rec}\rightarrow z=0)$ is the
density fluctuation growth factor from the recombination era at
$\bar{z}_{\rm rec}$ until the present epoch.  Therefore, instead
of Eq.(\ref{zto1070}), we have
\begin{equation}
\left(\frac{\delta T}{T}\right)\,\Big|_{z\sim {0}}\,= {\cal
K}\,\,\,\frac{\delta\rho}{\rho}\,\Big|_{{z}=\bar{z}_{\rm rec}}\,.
\label{ztorec}
\end{equation}

Using Eqs.(\ref{phizto1070}) and (\ref{first}), we have
\begin{equation}
\left(\frac{\delta\rho}{\rho}\right)\Big|_{z\sim
{0}}=\frac{F_1}{{\cal K}}\, {\cal D}({{z}_{\rm rec}}\rightarrow
z=0)\left(\frac{\delta T}{T}\right)\Big|_{{{z}_{\rm rec}}}
\label{transf1}
\end{equation}
and from Eqs.(\ref{phiztorec}) and (\ref{ztorec}),
\begin{equation}
\left(\frac{\delta\rho}{\rho}\right)\,\Big|_{z\sim
{0}}\,=\frac{F_1}{{\cal K}}\, {\cal D}(\bar{z}_{\rm
rec}\rightarrow z=0)\left(\frac{\delta
T}{T}\right)\Big|_{{{z}_{\rm rec}}} \,. \label{transf2}
\end{equation}

Equations (\ref{transf1}) and (\ref{transf2}) give the correction
factor $F_2$ due to the change in the value of the recombination
redshift,
\begin{equation}
F_2=\frac{{\cal D}(\bar{z}_{\rm rec}\rightarrow z=0)}{{\cal
D}({{z}_{\rm rec}}\rightarrow z=0)}\,.\label{f2}
\end{equation}
The growth of a perturbation in a matter-dominated Einstein-de
Sitter universe is $\delta\rho/\rho\propto a = (1+z)^{-1}$, where
$a$ is the cosmic scale factor \cite{coles}. Thus, the growth
factor $\cal D$ is
\[
{\cal D}\simeq (1+z)\,.
\]
We then find from Eq.(\ref{f2})
\begin{equation}
F_2\simeq \left(\frac{1+\bar{z}_{\rm rec}}{1+{z}_{\rm
rec}}\right)\,. \label{f2z}
\end{equation}

The temperature at ${{z}_{\rm rec}}$ in the standard model is
\begin{equation}
{T}\,({{z}_{\rm rec}})=T_0\,(1+{{z}_{\rm rec}})\,.
\end{equation}
In order for the temperature at the recombination epoch
$\bar{z}_{\rm rec}$, when the vacuum energy is decaying into CMB
photons,  to be $T ({z}_{\rm rec})$, we must have, from
Eq.(\ref{tempbeta}),
\begin{equation}
\bar{z}_{\rm rec}=(1+{{z}_{\rm rec}})^{1/(1-\beta)}-1\,.
\label{zdecay}
\end{equation}
From Eq.(\ref{f2z}), we then have
\begin{equation}
F_2\simeq (1+{{z}_{\rm rec}})^{{\beta}/{(1-\beta)}}\,.
\label{f2comp}
\end{equation}

The total factor $F$ is composed of $F_1$, due to the dilution of
the CMB as a result of vacuum energy decay, and $F_2$, due to the
change in the redshift of the recombination epoch. Assuming that
the effects described by $F^2_1$ and $F^2_2$ are independent and
that the total factor $F$ is the product of $F_1^2$ and $F_2^2$,
we have
\begin{equation}
F=F_1^2\,F_2^2\,. \label{factordecay}
\end{equation}

Thus, from Eqs.(\ref{f1beta}), (\ref{f2comp}) and
(\ref{factordecay}), the condition for the maximum value of $\beta
\in [0,1]$ is
\begin{equation}
{\beta}_{\rm max}= \alpha \left[ 1- \sqrt{ 1- \frac{ \ln{(F_{\rm
max})} } {2{\alpha}^2 \ln{(1+ {z}_{\rm rec}) } }    }\, \right]\,,
\label{betaeq}
\end{equation}
where
\begin{equation}
\alpha=1+\frac{ \ln{(F_{\rm max})} } {4 \ln{(1+ {z}_{\rm rec}) }
}\,. \label{alpha}
\end{equation}

As noted above, the maximum value of $F$ from observations is $F_{\rm max}\cong
1.1$. For ${z}_{\rm rec}\simeq 1100$, we find a very small maximum value of the
$\beta$ parameter, $\beta_{\rm max}\cong 3.4 \times 10^{-3}\,.$

\section{Conclusions}

We showed that the CMB and large galaxy survey data agreement put
strong limits on the rate of a possible decay of the vacuum energy
into CDM and into CMB photons.

When the vacuum energy decays into CDM, $\delta\rho/\rho$ is diluted and the density
fluctuation spectrum is amplified by a factor $F$ at the recombination era. The
$(\delta \rho/\rho)^2$ derived from the CMB and galaxy distribution data agree to
$\sim 10\%$, the maximum value for $F$ is then $F_{\rm max}\cong 1.1$.

We found that the decay of the vacuum energy into CDM as a scale factor power law
$\rho_{\Lambda}\propto (1+z)^{n}$, gives a maximum value for the exponent $n_{\rm
max}\cong 0.06$. For a parametrized vacuum decay into a CDM model with the form
$\rho_{\Lambda}(z,\nu)=\rho_{\Lambda}(z=0)+\rho_{\rm{c}}^0\,[\nu/(1-\nu)]\,
[\left(1+z\right)^{3(1-\nu)}-1]\, \,$, where $\rho_{\rm{c}}^0$ is the present
critical density, an upper limit on the $\nu$ parameter was found to be $\nu_{\rm
max}\cong 2.3\times 10^{-3}$.

We made a similar analysis for the possibility of the decay of the vacuum energy
into CMB photons. When photon creation due to the vacuum energy decay takes place,
the standard linear temperature dependence, $T(z)=T_0\,(1+z)$, where $T_0$ is the
present CMB temperature, is modified. We can place an upper limit on the $\beta$
parameter for the decay of the vacuum energy into CMB photons, parametrized by a
change in the CMB temperature at a given redshift $z$:
$\,\overline{T}(z)=T_0(1+z)^{1-\beta}\, $. We find that $\beta_{\rm max}\cong 3.4
\times 10^{-3}$.

Our results indicate that the rate of the decay of the vacuum energy into CDM and
CMB photons are extremely small. Since the results show that the vacuum energy can
only decay to a negligible extent into cold dark matter or CMB photons, we conclude
that if the vacuum energy is decaying, it is decaying, for example, into hot dark
matter (e.g., high energy neutrinos) or exotic matter (e.g., scalar fields), which
do not affect the $(\delta \rho/\rho)^2$ or the $\delta T/T$ CMB spectra.


\vskip 6mm
\noindent {\bf Acknowledgments.} R.O. thanks the Brazilian
agencies FAPESP (grant 00/06770-2) and CNPq (grant 300414/82-0)
for partial support. A.P. thanks FAPESP for financial support
(grants 03/04516-0 and 00/06770-2).



\end{document}